# Geometry dependent distribution of the supercurrent in YBa$_2$Cu$_3$O$_{7-x}$ films with patterned pinning landscape.

F. Laviano, D. Botta, A. Chiodoni, R. Gerbaldo, G. Ghigo, L. Gozzelino, B. Minetti, and E. Mezzetti

Istituto Nazionale per la Fisica della Materia, U.d.R. Torino-Politecnico, C.so Duca degli Abruzzi 24, 10129 Torino, Italy,

Istituto Nazionale di Fisica Nucleare, Sez. Torino, Via Pietro Giuria 1, 10125 Torino, Italy and

Department of Physics, Politecnico di Torino, C.so Duca degli Abruzzi 24, 10129 Torino, Italy

We created local pinning modulations in YBCO films by means of confined high energy heavy ion irradiation. The high energy of the ions allows us to introduce nanometric size defects with a well defined anisotropy. The dose was chosen in such a way to reduce the local critical current of the irradiated area. We used a quantitative magneto-optical analysis to measure the magnetic field vector and the supercurrent for each point of the whole sample surface. The basic geometry of a rectangular region inside strip-shaped samples was considered in order to investigate in detail the effect of the orientation of planar boundaries with respect to the supercurrent flow direction. Here we present the two complementary orientations of the modulated region, i.e., perpendicular and parallel to the main supercurrent flow. The comparison of the magnetic field and supercurrent distributions shows deep differences between the two configurations. In particular, the enhanced vortex diffusion, observed for the perpendicular case, was not found in the parallel configuration. In a such case, unexpected vortex bundle jumps and a Meissner volume compression are clearly observed after the vortices enter the irradiated region.

74.60.Ge, 74.60.-w, 74.76.Bz, 74.80.-g

## Introduction

The local modulation of the superconducting properties allows studying the fundamental interactions of the vortex matter[1] as well as its phase transitions in confined geometries[2,3] and it can be exploited for designing superconducting devices.[4] The electrodynamics of high temperature superconducting (HTSC) oxides is affected by crystal lattice modifications over several length scales, from the nanometric size of the vortex pinning centers to the interaction length of the screening currents.
In the transverse geometry, i.e., in thin specimens with perpendicular external magnetic field,[5,6] the two-dimensional (2D) distribution of the supercurrents results in a nonlocal interaction between the magnetic flux and screening currents on the length scale of several $\lambda$ (where $\lambda$ is the effective London penetration depth),[7] while the screening current distribution itself is governed by geometrical constraints over the whole size of the sample.[8]
K.W. Kwok et al.[9] have recently demonstrated the possibility to periodically pattern the vortex pinning landscape on the micron scale by means of the high-energy heavy-ion (HEHI) irradiation.[10] The nanometric structures created by the HEHI irradiation are pinning sites correlated along the beam direction[11,12] and in fact they are called "columnar defects" (CDs). This correlation leads to a well-defined anisotropy of the pinning properties depending on the direction of the local magnetic field.[13,14,15]

Thus a micron-scale pattern (micropattern) created by confined HEHI irradiation is expected to strongly influence both the local pinning properties and the whole supercurrent distribution in HTSC films. Moreover, due to the strongly nonlinear relation between the local supercurrent and electric field, the geometry of the micropatterns is a crucial parameter.[16]

Recent magneto-optical measurements directly visualized the vortex distribution around planar defects perpendicularly aligned with respect to the supercurrents.[17] The enhanced magnetic flux diffusion at defect interfaces was well described by taking into account the electric field focusing at such planar boundaries.[18] Also the vortex distribution observed in low-angle grain boundaries was related to the enhancement of the local electric field.[19] However, the electrodynamics of the whole sample could drastically depend on the orientation of planar boundaries with respect to the supercurrent due to the strong nonlinear *E-J* relation and the nonlocal interaction between vortices and screening currents in thin superconductors.

In order to study the influence of the shape and position of micron-scale pinning modulations on the superconducting electrodynamics, we used confined 0.25 GeV Au irradiation to tailor the pinning properties of rectangular regions in $YBa_2Cu_3O_{7-x}$ (YBCO) strips. The rectangular regions, here addressed as microchannels, were positioned either perpendicular (transverse) or parallel (longitudinal) to the supercurrent flow, supposed to be one-dimensional in strip-shaped homogeneous samples. We employed the quantitative magneto-optics (QMO)[20,21] to reconstruct the magnetic induction vector at the sample surface and the local supercurrents.

Very different behaviours are displayed when neither the strength of the pinning modulation nor the sample shape but only the microchannel orientation with respect to the supercurrent flow was changed. In both cases, as expected, the micropatterns influence the superconducting properties well beyond the irradiated area. A completely novel, unexpected, phenomenology is observed for the parallel microchannel.

The paper is organized as follows. Section 2 is devoted to the experimental details. In Section 3 the results for the virgin sample and for the patterned samples, with a transverse or longitudinal microchannel, are separately reported and discussed. In Section 4 the different phenomenologies, as observed for the two micropattern geometries, are analysed and summarized.

**Experimental details**

The YBCO films (c-axis oriented, $T_C \sim 89$ K, $\Delta T_C \sim 0.5$ K, thickness of 300 nm) were grown by thermal co-evaporation on yttria stabilized zirconia (YSZ) substrates, with a buffer layer of $CeO_2$ (typical thickness of 40 nm)[22]. Two long strips (1.3 x 0.3mm$^2$) were obtained by standard photolithography and chemical wet etching.

The strips were irradiated at high fluence (about $5 \cdot 10^{10}$ cm$^{-2}$, dose equivalent field $B_\Phi \sim 1$ T), with 0.25 GeV Au ions directed perpendicular to the film plane (z axis).To create the microchannels, the HEHI beam was confined by means of a slit into a stainless steel mask (laser cut 2x0.07 mm$^2$). One of the strips was irradiated with the slit perpendicular to its long side, the other with the slit in the complementary configuration, i.e., parallel to the long side. In the latter case, the irradiation area was positioned in order to influence only one half (direction) of the main supercurrent flow.

The HEHI distribution through the mask was checked both by MonteCarlo simulations and by the measurement of the darkening level of polyester films irradiated in the same way as the samples. This procedure was used to quantitatively measure the mean irradiation dose, see Figure 1.

We point out that HEHIs pass through the YBCO film and the buffer layer, where they induce columnar defects,[10,11] and afterwards they are implanted at less than 13 μm (calculated by SRIM2003©) into the YSZ substrate (500 μm thick). The damage caused by ion collisions (e.g., bright areas shown in the transmission electron microscopy (TEM) measurements of Figure 2a and Figure 2b) and the ion implantation induce local expansion of the crystal lattice constants of the substrate. The boundary between pristine and irradiated regions exhibits a small upward step, as depicted by the atomic force microscopy (AFM) measurements (Figure 2c). This suggests that the substrate damage propagated in the whole hetero-structure. In our case, the damage brought by the selected irradiation dose depresses the local critical current, allowing a deep pinning modulation and thus a clear observation of the presented phenomenologies.

Magneto-optics was used to locally measure the $B_z$ magnetic induction component over the samples, and then to evaluate the local current density and the other magnetic field components by the inversion of the Biot-Savart law.[20] The reconstruction of the actual current density distribution ($J(x,y)$ integrated over the thickness) is a model independent problem for thin superconductors in transverse geometry.[5,20] The local $B_z$ values, obtained by nonlinear calibration of the measured magneto-optical response, are corrected taking into account the coupling between the indicator film and the magnetic field components parallel to the indicator plane.[21,23] Moreover, the evaluation of all the magnetic induction components, ($B_x(x,y)$, $B_y(x,y)$, $B_z(x,y)$), directly gives the tilt angle of the magnetic field lines at the sample surface, $\theta(x,y)$.[15] QMO measurements were carried out at low temperatures (close to 4.2 K), after zero field cooling (ZFC). Then, magnetic fields of increasing intensity were applied (3 mT steps) perpendicular to the films and the corresponding frames were recorded (3s delay for each frame after the field set up).

## Results and Discussion

### a) Pristine strip

Pristine samples show high homogeneity of the intrinsic pinning landscape as demonstrated by the continuous and unperturbed supercurrent distribution (Figure 3a). A homogeneous critical current due to the vortex density gradient is established in the part of the sample where the vortices remain pinned, while in the whole central Meissner volume the screening supercurrent flows according to the screened volume geometry.[8] In figure 3b, we present a zoom on the region where the current density bends and forms a so-called discontinuity lines (d-lines).[24] It is worth noting that regions where the supercurrent changes only its direction corresponds to $d^+$-lines (indicated by the dashed line in Figure 3b), whereas if the current density changes its modulus, $d^-$-lines are present, e.g., on the sample edges.[25] With increasing external field, $d^-$-lines are characterized by enhanced dissipation, because vortices pile up at their locations, and $d^+$-lines are expected to repulse vortices, thus being location of minima of the electric field.[26]

### b) Perpendicular microchannel

The HEHI irradiation with the selected fluence causes a depletion of the pinning strength in our heterostructure, due to substrate damage at high irradiation doses, as discussed above. If the microchannel is perpendicular to the supercurrent flow and the pinning strength is reduced, the critical current must bend at the microchannel interfaces in order to be a continuos vector field. The location of the nanostructured region is clearly visible in the $B_z$ distribution, because of the increased flux density around the defected zone,[17,18,19] see Figure 4a. The magnetic induction exhibits peaks both at the microchannel boundaries and at the sample edges, since these are all indeed $d^-$-lines.[25] In addition to the critical current bending, the triple $d^+$-line at each microchannel side implies a Meissner supercurrent path running through the irradiated part and joining the flux free space so that a continuous Meissner volume extends across the whole sample. The triple $d^+$-line was also found in YBCO films grown on low-angle bi-crystal substrates.[27]

The main supercurrent flow is actually affected on a larger length scale with respect to the width of the micropattern, as shown by the current density profile in Figure 4b. Absolute minima of the supercurrent modulus are due to the weak-link property of the irradiated/pristine interfaces[28] and they are correlated to the characteristic HEHI collision profiles found both in simulations and in irradiated calibration sheets (see Figure 1). The enhanced penetration of the vortices along the interface region is expected for the presence of a local ridge of the electric field, which occurs at the boundary between different critical current regions ($d^-$-lines).[16,17,18,19,25] Inside the bulk of the microchannel (Figure 5) vortices are readily nucleated and diffuse deeper with respect to the pristine banks of the strip also due to the suppression of surface barriers by CD creation at the sample edges.[29]

The anisotropy of CDs is revealed by QMO, because the vortex curvature increases moving towards the central Meissner zone and thus the pinning efficiency of CDs decreases (see Figure 6).[13,15] The Meissner state, although confined in a thinner and thinner region as the external magnetic field is raised, holds on across the whole sample for the whole range of the applied magnetic field ($0 \div 0.2$ T at T = 5 K).

**c) Parallel microchannel**

The second microchannel exhibits a rich and quite unexpected phenomenology. The magnetic induction pattern, presented in Figure 7a, breaks up into three regions bounded by the $d^+$-lines starting from the corners of the sample. For top and bottom parts of the sample, we observe that the vortex arrangement is similar to the previous one, because the supercurrent direction is there perpendicular to the irradiated/pristine interfaces. On the contrary, where the critical current flows parallel to the channel boundary, i.e., parallel to the longest edge (longitudinal part) of the micropattern, a new kind of electromagnetic response is reported (see Figure 7b).

The behaviour observed during vortex diffusion after ZFC can be summarized in two phases. In the former, the longitudinal part displays the same values and distribution of the current density, as before irradiation, until the first vortices approach it. This suggests that the Meissner supercurrent are unaffected by the given density of CDs and by the structural stress from the damaged substrate.

Then vortices, moving from the right sample edge into pristine region, reach the interface, Figure 8a, and start to diffuse into the micropattern in an unexpected way, Figure 8b and 8c. This second phase is characterized by the appearance of vortex bundles, which jump over the longitudinal channel interface as clearly observed by the difference of QMO frames with successive external applied fields, shown in Figure 9. The bundles are nucleated after vortices run through preferential paths, indicated by arrows in Figure 9. A further comparison among the MO images indicates that vortex jumps are triggered from the points where vortices overtake the envelope of the flux front (see Figures 7 and 8), i.e., from loci of electric field focusing.[26] Moreover, the bundles jump over a domain of $B_z$ minima near the interface region, which looks like a $d^+$-line. The presence of this extrinsic $d^+$-line is confirmed by the absence of any discontinuity in the current density profiles and by the appearance of $B_z$ maxima (Figure 10), when decreasing the external field.[25] It follows that flux bundles, whose typical size spreads from very few vortices (1-10) to over 1000 flux quanta, surprisingly cross this $d^+$-line to reach the channel bulk. This behaviour is puzzling because $d^+$-line are expected to be never crossed by vortices,[25] during external field increasing. Moreover, in contrast to the behaviours in the top and bottom parts of the same microchannel interfaces, no enhanced vortex diffusion is found.

Another striking feature is observed in this second phase of the vortex diffusion process: when the vortices enter the microchannel, the Meissner volume shape is modified inside the whole sample. Consequently the central $d^+$-line, where the main supercurrent loop collapses, is displaced, as it is observed in the profile comparisons between pristine and irradiated samples of Figure 8. Such deformation of the Meissner volume shape is also indicated by the flux line curvature discontinuity when vortices cross the microchannel. Following the sharp modulation of the critical current into the irradiated region, $\theta(x,y)$ displays two discontinuities at the interfaces (see profiles of Figure 11). Therefore, the Meissner volume (ideally of ellipsoidal section) is subject to two subsequent deformations: at first, its right part exhibits a decreased curvature, but reduces also its lateral extension (vortices diffuse deeper in the irradiated region with respect to the pristine part) and after, more curved flux lines actually squeeze up the Meissner volume and the displacement of the $d^+$-line becomes very appreciable (Figure 12). Since the Meissner volume deformation implies a redistribution of the supercurrent in the whole sample and occurs only when vortices reach the microchannel, we ascribed the observed phenomenology to a pronounced nonlocal interaction between vortices and screening currents for the transverse geometry.

## Conclusions

We created micrometric scale patterns in YBCO thin strips by means of confined HEHI irradiation, with the aim of inducing and studying local supercurrent modulations. Firstly, we compared the two micropattern orientations (parallel or perpendicular) with respect to the nonlinear supercurrent flow direction. The QMO analysis showed huge differences in the vortex behaviour between the two configurations. Secondly, by taking into account the modulation attributed to the interaction between anisotropic CDs and curved magnetic flux lines, we analysed the influence of the planar channel boundaries on the shape of the Meissner volume embedded in the bulk of both the microchannels. For supercurrents crossing the microchannel, we observed an expected strong enhancement of the magnetic flux diffusion due to the dominating role of the electric field focusing at the interfaces ($d^-$-lines ). For the parallel configuration, a pronounced nonlocal vortex diffusion (due to the thin film geometry) is observed when vortices pass over the microchannel interface and cause an abrupt rearrangement of the coherent Meissner state in the whole sample. This observation demonstrates that the Meissner volume does not retain a constant shape during vortex penetration inside inhomogeneous superconductors. Moreover, peculiar vortex bundles, generated from the main vortex front, are forced to jump at 'hot-spot' locations near the interface (maxima of the electric field). The jumps occur across a visible d+-line. Such a line was never observed to be crossed by vortices when the external magnetic field is increased.

In conclusion, the observed phenomenologies show that the orientation and position of micropattern interfaces, with respect to the supercurrent flow direction, play a fundamental role in confined geometries and therefore have to be taken into account or even exploited for superconducting devices.

**Acknowledgments**

YBCO samples was kindly supplied by the joined team "Edison-Europa Metalli-IMEM/CNR" under a partially funded CNR project (L.95/95). We acknowledge the support of the Istituto Nazionale di Fisica Nucleare and the Istituto Nazionale per la Fisica della Materia for partial funding. The ESF Vortex Program is gratefully acknowledged.

**Figure 1. MonteCarlo simulation (line) and measured polyester sheet darkening level (points) of HEHI irradiation through a metallic mask with a microscopic slit (about 70 µm wide). The two peaks are due to the HEHI scattered by the slit edges of the metal mask.**

**Figure 2. a) High resolution in-plane view of the YBCO film with one columnar defect present (bright spot). b) High resolution in-plane view of the single crystal YSZ substrate by TEM. Two defects caused by the HEHI collisions are visible. c) AFM topography of the irradiated/pristine interface region showing an increased mean thickness of the film, in the irradiated area, of about 20±5 nm.**

**Figure 3. a) Magnetic field ($B_z(x,y)$) and current density modulus ($|J(x,y)|$) distributions for a pristine sample at T= 4.13 K and $\mu_0 H_{app}$= 117.3 mT. b) Zoom**

view in the dotted window, on the upper left corner of the $B_z(x,y)$ map, and current density streamlines superimposed. The bisector of the corner and the $d^+$-line domain, bounded by two straight lines, are also depicted.

Figure 4. Strip with perpendicular microchannel. a) $B_z(x,y)$ and $|J(x,y)|$ at T= 5.65 K and $\mu_0 H_{app}$= 88 mT. The dashed lines indicate the triple $d^+$-line. b) Current density modulus profile across the microchannel region (dotted lines in $|J(x,y)|$ map).

Figure 5. Zoom view of the perpendicular microchannel ($B_z(x,y)$ with current density streamlines superimposed) and corresponding current density profiles on the irradiated and the pristine locations, indicated by the straight lines across the strip in Figure 4a (*A* and *A'*, respectively). a) T= 5.65 K and $\mu_0 H_{app}$= 29.5 mT; b) T= 5.65 K and $\mu_0 H_{app}$= 58.8 mT; c) T= 5.65 K and $\mu_0 H_{app}$= 88 mT.

Figure 6. $|J(\varphi)|$ curves (T= 5.65 K, $\mu_0 H_{app}$= 88 mT), in the vortex-state region ($B_z \neq 0$) both in pristine and in irradiated parts (profiles along *A* and *A'*). There is evidence for strong critical current dependence on $\varphi$ due to the columnar defect anisotropy. Into pristine region, critical current is almost constant up to large $\varphi$. $J_{max}$ is the maximum of the supercurrent into the critical state region.

Figure 7. $B_z(x,y)$ and $|J(x,y)|$ distributions for the sample with parallel microdefect at T= 4.23 K and $\mu_0 H_{app}$= 88 mT. The area enclosed in the dashed box (in the $B_z(x,y)$ map) is reported with larger magnification in Figure 8.

**Figure 8.** Zoom view of $B_z(x,y)$ map with current density streamlines for the longitudinal microchannel. The data plotted on the bottom are taken on the same location (dashed line in Figure 8a), before and after the microdefect creation. a) T= 4.23 K and $\mu_0 H_{app}$= 58.8 mT; b) T= 4.23 K and $\mu_0 H_{app}$= 73.4 mT; c) T= 4.23 K and $\mu_0 H_{app}$= 88 mT.

**Figure 9.** Differential image of subsequent MO frames at $\mu_0 H_{app}$= 58.8 mT and at $\mu_0 H_{app}$= 55.8 mT, T= 4.23 K. The arrows point to the locations where straight dark paths with almost unmodified vortex density indicate the corresponding vortex bundle jump into the channel. The saw-tooth like signal inside the Meissner area is due to perpendicular magnetic domains of the indicator film.

**Figure 10.** $B_z(x,y)$ distribution at $\mu_0 H_{app}$= 58.8 mT after applying the maximum magnetic field of 180 mT (decreasing field branch of the magnetization cycle), at T= 4.23 K. Note that the sample was not fully penetrated with the maximum applied field, thus there is still a central region inside the superconductor that was never reached by vortices.

**Figure 11.** Profiles of $q(x,y)$ across the longitudinal microchannel at $\mu_0 H_{app}$= 58.8 mT and at $\mu_0 H_{app}$= 88 mT. Discontinuities of the tilt angle $q$ occur at both interfaces.

**Figure 12.** Displacement of the central $d^+$-line with the longitudinal microchannel, during the virgin magnetization cycle. These points correspond to

**places where the screening current value crosses the zero. The error bars represent an uncertainty of ±1 pixel (1.65 μm).**

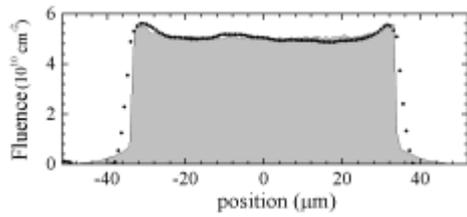

Figure 1.

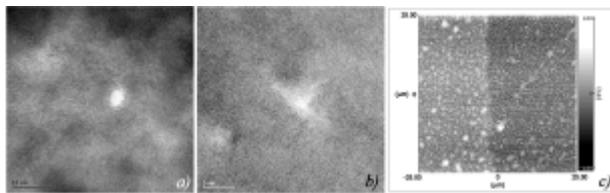

Figure 2.

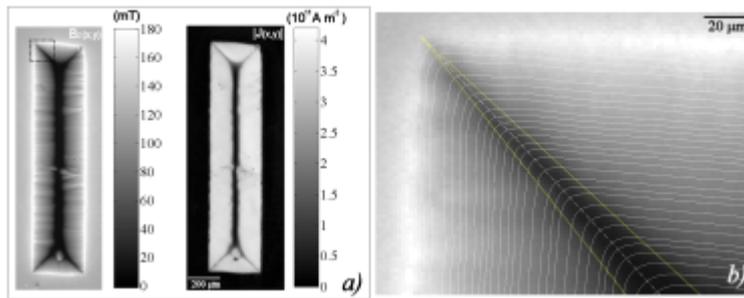

Figure 3.

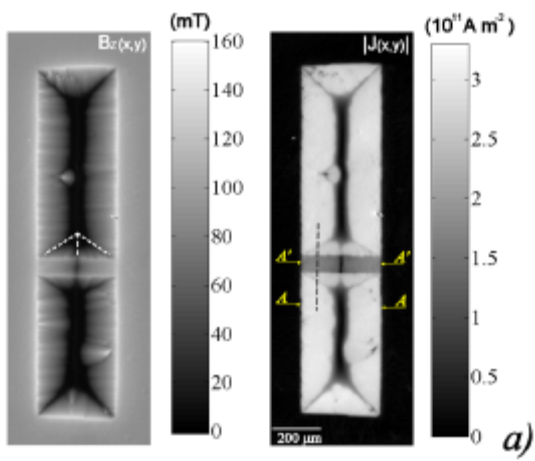

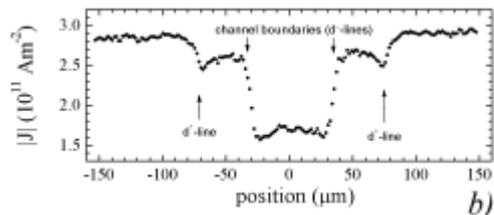

b) Figure 4.

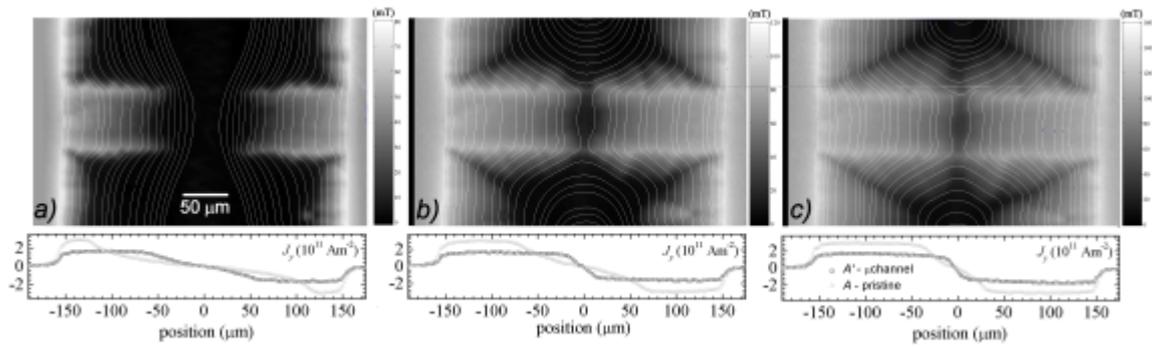

Figure 5.

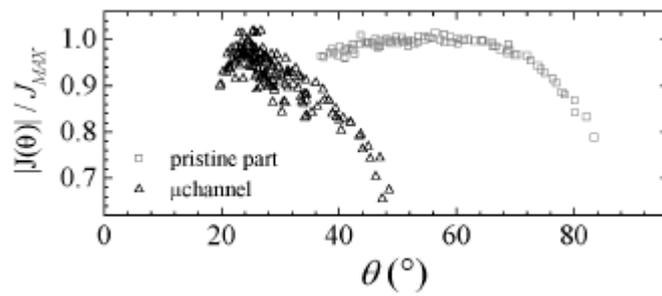

Figure 6.

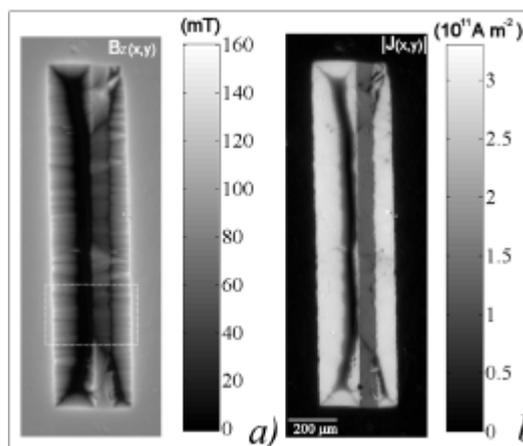

b) Figure 7.

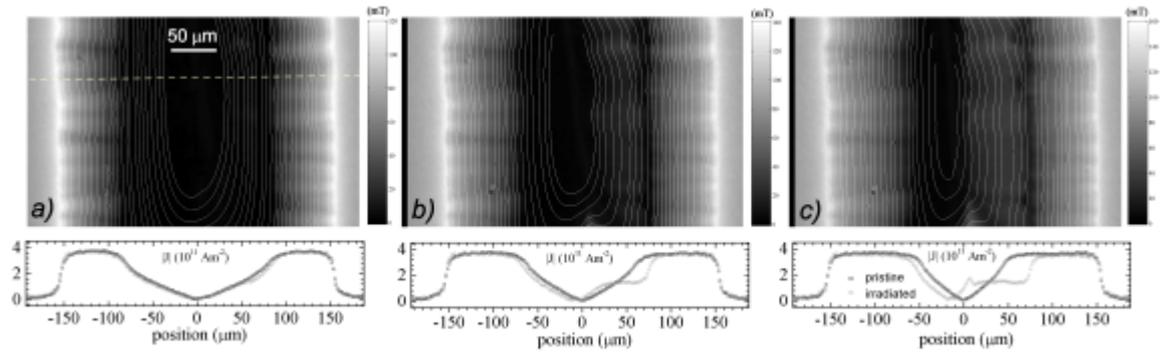

Figure 8.

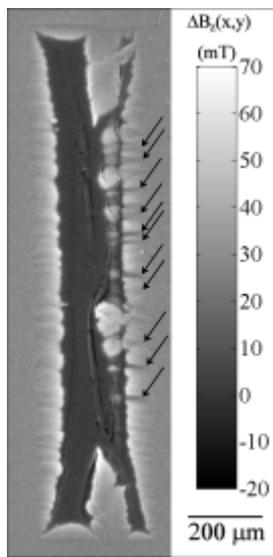

Figure 9.

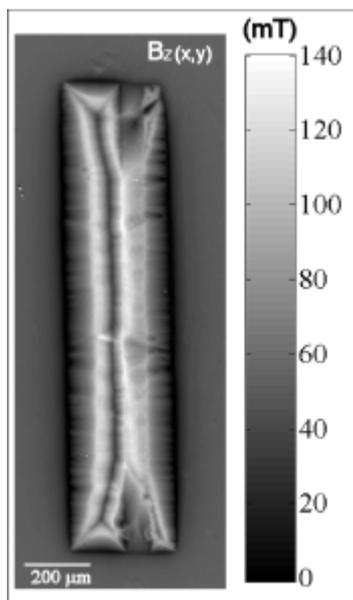

Figure 10.

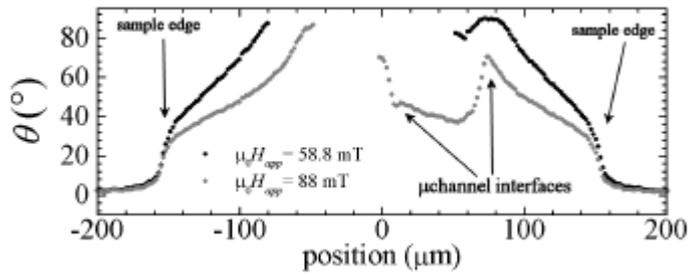

Figure 11.

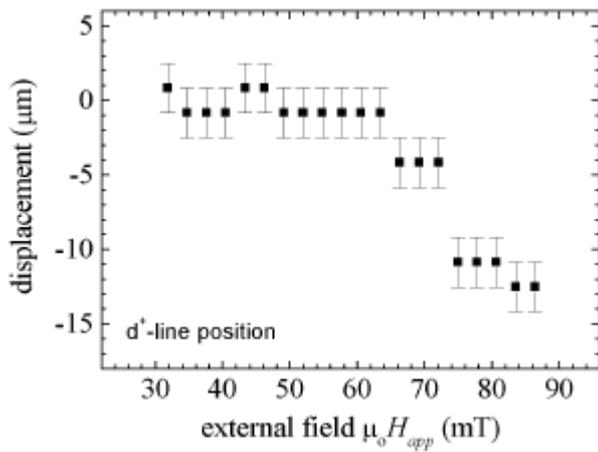

Figure 12.